\documentclass[sigconf,nonacm]{acmart}

\usepackage{framed}
\definecolor{shadecolor}{gray}{0.94}

\title{Bespoke: Generating MOOC-Quality Industry-Personalized Lecture Videos at Scale}

\author{Romain Puech}
\email{puech@mit.edu}
\author{Dewang Kumar Agarwal}
\author{Antonio Santamaría Escobar}
\affiliation{%
  \institution{Massachusetts Institute of Technology}
  \city{Cambridge}
  \state{Massachusetts}
  \country{United States}
}

\author{Dimitris Bertsimas}
\affiliation{%
  \institution{Sloan School of Management, Massachusetts Institute of Technology}
  \city{Cambridge}
  \state{Massachusetts}
  \country{United States}
}

\renewcommand{\shortauthors}{Puech et al.}

\begin{abstract}
Lifelong learners from different industries often watch the same recorded lecture, even when they will apply the material in different workplaces. We present Bespoke, a system that takes the transcript of an existing lecture and generates a new video customized to a target industry and duration, with new slides, narration, and charts. From 31 graduate lectures in analytics, machine learning, and optimization, we generated 209 videos for these three industries, plus a generic-audience version of each lecture. Twenty-five experts in the corresponding domains rated 92 videos on a five-point rubric. They judged 87\% at or above the rubric midpoint corresponding to ``a standard MOOC lecture's quality'' (mean overall score $3.42$ out of $5$; 48\% scored $4$ or above). Overall quality was similar across industries, durations, and a 21-lecture held-out set unused while developing the system, at about \$0.22 API cost per minute of video.
\end{abstract}


\begin{document}

\begin{teaserfigure}
  \includegraphics[width=\textwidth]{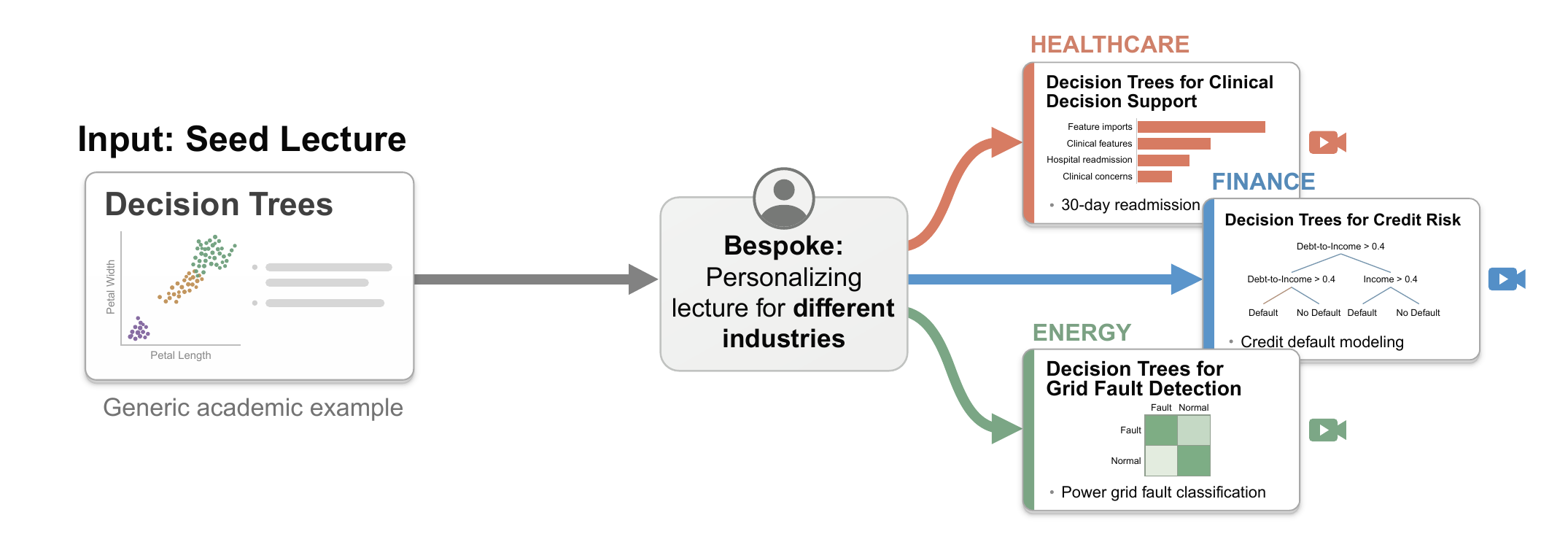}
  \caption{Bespoke generates a new lecture from a seed transcript for a stated industry or a generic audience.}
  \Description{A seed decision-tree lecture with a generic iris example is regenerated for healthcare (readmission), finance (credit risk), and energy (grid faults).}
  \label{fig:concept}
\end{teaserfigure}

\maketitle

\section{Introduction}

Asynchronous lecture videos are a primary medium for professional upskilling in MOOCs and corporate learning platforms~\cite{jordan2015,kizilcec2013,kizilcec2017}.
A recorded lecture typically uses one set of examples and one level of technical depth for every viewer.
That design fits poorly when the viewers are working professionals who will apply the same method in different industries.
A healthcare analyst studying decision trees needs hospital readmission prediction and explainability; a financial engineer cares about credit-default features; an energy-sector manager needs the gist of grid fault classification without technical depth.
When the instructor uses a generic academic example, such as classifying iris species, the lecture stays far from the workplace where each of those viewers will use it.

Rewriting slides and re-recording narration by hand for every industry does not scale.
Recent systems personalize lecture video while leaving the original recording in place.
VIVID helps instructors turn a monologue lecture into vicarious dialogue~\cite{choi2024}.
PedaCo-Gen, an extended abstract, keeps the educator in a co-authoring loop for text-to-video~\cite{baek2026}.
SAM~\cite{bodonhelyi2025} and Generative Lecture add interactive Q\&A on top of an existing video~\cite{jo2025}.
Adaptive platforms recommend which existing item a learner sees~\cite{kizilcec2017}, and PAGE personalizes educational text to a student profile~\cite{lim2025}.
Guo, Kim, and Rubin~\cite{guo2014} showed that short, purpose-made videos engage better than long classroom recordings, which still leaves an instructor to write each short video.

Those systems either recommend an existing recording or add interaction on top of one.
Bespoke takes a third route: it regenerates the lecture, with new slides, narration, and charts, for a stated professional audience.

We present Bespoke, a five-stage system that takes the transcript of an existing lecture, a target industry, and a target duration, and generates a new lecture video: new slides, narration, and charts, with examples drawn from that industry.
The same technical topic can also be generated for a generic audience, so industry customization can be compared with a version that keeps academic examples.
Bespoke writes audience-specific learning objectives, plans the lecture around a retrieved domain paper, writes narration, builds slides and charts, and reveals slide elements as they are spoken.
Generation used Anthropic's Claude Sonnet~4.6 API and OpenAI's GPT-5.2 API, with Sonnet generating and GPT-5.2 validating, in a bounded refine loop.
Objectives are written before content, following the learning-sciences tradition of backward design~\cite{wiggins2005}, and Mayer's multimedia principles~\cite{mayer2009} constrain generation.
We generated over 200 videos from 31 graduate lectures, including 21 unused while developing the system, at about \$0.22 API cost per minute of video.

We evaluate Bespoke with 25 reviewers who have teaching or research experience in healthcare, finance, energy, and machine learning.
They rated a 92-video sample drawn from that corpus on a five-point rubric whose overall item $G$ uses ``comparable to a standard MOOC lecture'' as the midpoint anchor (full rubric in Appendix~\ref{app:rubric}).

This motivates the following question:

\noindent\textbf{RQ1.} Can Bespoke produce industry-personalized lecture videos that domain experts judge adequate, at scale?

We report two contributions:

\begin{enumerate}
\item Bespoke, a pipeline that regenerates a complete lecture video for a stated professional audience from a seed transcript, compiling slides, narration, and charts.
\item An expert evaluation of 92 videos on a four-dimension rubric reproduced in Appendix~\ref{app:rubric}. Reviewers judged 87\% at or above the MOOC-comparable rubric level (mean $G=3.42$ out of $5$), with similar overall scores across industries, durations, and the 21-lecture held-out set, at about \$0.22 API cost per minute. Reviewers' free-text comments, together with how rubric dimensions associate with overall score, identify voice, slide timing, and layout as the main remaining issues.
\end{enumerate}

The next section reviews related work on lecture video and personalization.
We then describe the system, the expert study, and the results.

\section{Related Work}\label{sec:related}

We situate Bespoke in four strands of HCI and learning-at-scale research: lecture video as a designed medium, human--AI authoring of instructional video, personalization of what a learner sees, and evaluation of generated teaching artifacts.

\subsection{Lecture Video as an HCI Medium}

HCI and Learning at Scale treat lecture video as a designed object.
Guo, Kim, and Rubin analyzed millions of edX watching sessions and showed that length, talking-head versus classroom capture, and studio style predict engagement~\cite{guo2014}.
Shorter, purpose-made segments outperform long classroom recordings.
That finding motivates writing a new lecture for a new audience and a new length.

VIVID treats lecture form as an authoring problem~\cite{choi2024}.
Choi et al.\ help instructors turn a monologue lecture into vicarious dialogue while the original video stays on screen and the instructor remains the author.
Bespoke keeps the monologue lecture as the output genre and changes the audience: examples, framing, and depth are rewritten, and new slides, narration, and charts are generated from the transcript.

\subsection{Human--AI Authoring of Instructional Video}

A second line of work keeps the educator in a co-authoring loop.
VIVID's three-stage workflow (generate, compare, refine) lets instructors select and edit LLM-proposed dialogues~\cite{choi2024}.
PedaCo-Gen, an extended abstract, uses Mayer's Cognitive Theory of Multimedia Learning as generation constraints and an intermediate video blueprint that educators review before text-to-video synthesis~\cite{baek2026}.
In a study with 23 education experts, the authors report higher pedagogical quality than unconstrained baselines.

Bespoke uses the same Mayer principles at generation time and the same style of expert rating.
It generates a complete lecture for a stated professional audience without an instructor in the loop at generation time, and it compiles HTML slides, code-generated charts, and narrated audio.
Text-to-video models optimize visual realism~\cite{brooks2024,deepmind2025} and are expensive at lecture length; they also have trouble with equations and deterministic diagrams.

Adjacent multi-agent systems generate pixel video for storytelling (GenMAC~\cite{huang2026}; Hollywood Town~\cite{wei2025}; MAViS~\cite{wang2026}).
Slide-generation tools produce structured decks from documents or instructions (AutoPresent~\cite{ge2025}; PPTAgent~\cite{zheng2025}; Auto-Slides~\cite{yang2025}).
LASEV~\cite{yan2026} is the closest compile-to-video system: specialized agents emit code and text that assemble into an educational clip.
Its template-based design is built for narrow task types (short worked examples in math or language) and clips of about one minute.
Bespoke generates open-ended graduate lectures of 5 to 45 minutes from an existing human lecture, across healthcare, finance, energy, and a generic professional audience, including a 21-lecture held-out set unused during development, and we evaluate those videos with domain-matched experts.

\subsection{Personalizing What the Learner Sees}

SAM (Study with AI Mentor) overlays a context-aware LLM chat on an existing lecture video so students can ask about formulas, slides, and images in real time~\cite{bodonhelyi2025}.
Generative Lecture inserts an AI instructor clone that answers questions during playback, leaving the original lecture intact~\cite{jo2025}.
In both cases the student initiates the adaptation and the seed video stays.
Adaptive MOOC platforms recommend which existing video or module a learner sees, based on diagnostics or engagement~\cite{kizilcec2017}.

PAGE personalizes notes, explanations, and worked examples to a student profile inside an intelligent tutoring platform~\cite{lim2025}.
A semester-long study reports gains in learning and engagement.
Stavrinou et al.~\cite{stavrinou2025} generate short social-media reels from existing lecture video; the unit is a clip, and there is no industry retargeting.
Bespoke's unit of personalization is one full lecture for one professional audience (or a generic-audience version of the same lecture).

\subsection{Evaluating Generated Teaching Artifacts}

VIVID and PedaCo-Gen, like other CHI systems that generate instructional media, evaluate with instructors or education experts rating pedagogical quality~\cite{choi2024,baek2026}.
We follow that mode: 25 domain-matched reviewers scored generated lecture videos on a four-dimension rubric.

VIVID and PedaCo-Gen keep instructors in a co-authoring loop; SAM and Generative Lecture add Q\&A on top of the original video; adaptive platforms recommend existing items.
Bespoke generates a new lecture, with new slides, narration, and charts, for a stated professional audience, and we evaluate those videos with 25 domain-matched experts.

\section{System Design}\label{sec:design}

Our goal is to generate a personalized lecture video from a seed lecture and a target learner profile.

\subsection{Modeling Choices}

\textbf{Regenerate from a seed transcript.}
Every lecture starts from a human instructor's seed transcript $S$.
The transcript defines the topic and scope and supplies the facts the new lecture may use.
We take speech text because the content that matters is spoken or inferable from speech.
Replacing one plot or example in the original video tends to break neighboring slides and narration; generating a new lecture from $S$ keeps the whole artifact consistent.

\textbf{Compile slides, charts, and narration.}
Text-to-video and text-to-image models~\cite{brooks2024,deepmind2025} optimize visual realism and have trouble with equations, diagrams, and stepwise argument.
Bespoke writes HTML slides with rendered math, code that draws charts and diagrams, and a narration transcript, then compiles those artifacts into a video.
Code and markup are native LLM training representations, which helps syntactic validity and leaves files an instructor can edit.
The videos have no talking-head instructor; Mayer's image principle reports that an on-screen instructor image does not improve learning~\cite{mayer2009}.

\textbf{Inputs and outputs.}
The pipeline takes a seed transcript $S$, a learner profile $\pi$ (healthcare, finance, energy, or a generic audience), a target duration $d$ (medium, 5--15 minutes, or long, matching the seed up to about 45 minutes), and optionally instructor-supplied learning objectives or extra instructions.
It outputs a lecture video with industry-adapted slides, synthesized narration, and charts.

\subsection{Architecture}

\textbf{Fixed stage order.}
Bespoke is a directed acyclic graph with a fixed stage order.
The only branching is a bounded generate--validate--refine loop that always terminates.
A fixed workflow gives predictable cost and latency because the task already decomposes into sequential steps.
All videos in this evaluation were generated without instructor intervention at generation time.

\begin{figure*}[t]
  \centering
  \includegraphics[width=\linewidth]{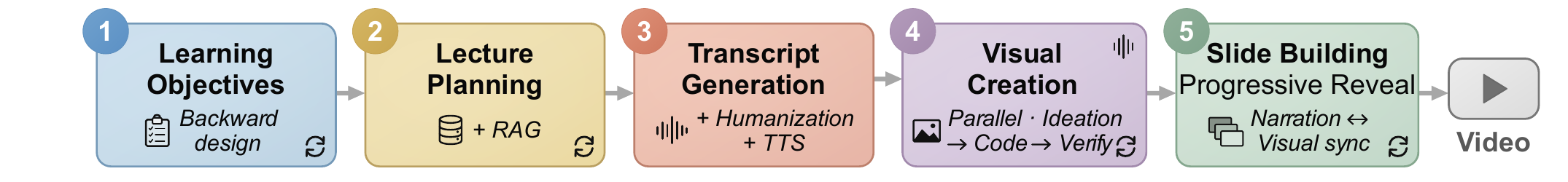}
  \Description{Five stages: learning objectives, lecture planning with retrieved-domain grounding, narration, parallel visual creation, and progressive reveal, ending in a compiled video.}
  \caption{Five stages: objectives, retrieved-domain planning, narration, parallel visuals, and progressive reveal.}
  \label{fig:pipeline}
\end{figure*}

\textbf{Five stages.}
The pipeline writes audience-specific learning objectives, plans the lecture around a retrieved domain paper, writes narration, builds slides and charts in parallel, and reveals slide elements as they are spoken (Figures~\ref{fig:concept} and~\ref{fig:pipeline}).
Audio is then synthesized and assembled with the slides.

\textbf{Generate, validate, refine.}
Checking a draft against a specification is easier than getting the draft right on the first try~\cite{madaan2023}.
Allocating extra inference to revision also tends to improve quality~\cite{snell2025}.
A generator produces an artifact; a validator from a different model family returns a structured critique; the generator revises.
Using Anthropic's Claude Sonnet~4.6 API to generate and OpenAI's GPT-5.2 API to validate reduces self-enhancement bias~\cite{panickssery2024}.
A bound of three refinement iterations keeps cost finite.

\textbf{Programmatic checks where possible.}
Duration arithmetic, schema checks, and detection of missing images or overflow run in ordinary code.
The LLM is reserved for semantic judgments: audience fit, visual intent, and conversational register.

\textbf{Backward design and multimedia principles.}
Following backward design~\cite{wiggins2005}, objectives are written before content.
The revised Bloom taxonomy's action verbs~\cite{anderson2001} constrain how those objectives are phrased.
Mayer's multimedia principles~\cite{mayer2009} of coherence, segmenting, redundancy, signaling, spatial contiguity, and conversational personalization are implemented as generation prompts and as named validation criteria.

\section{The Bespoke Pipeline}\label{sec:pipeline}

We walk one running example through the five stages: a seed lecture on decision trees that classifies iris species, rewritten for clinical data scientists at a regional hospital.

\subsection{Stage 1: Audience-Specific Learning Objectives}

The first stage commits to objectives before any content is planned.

In the running example, the seed's implicit goal is ``understand how decision trees work.''
Stage~1 might produce:
\begin{quote}
By the end of this lecture, learners will be able to evaluate the trade-off between interpretability and accuracy when presenting diagnostic recommendations to clinical staff, and to identify when a decision tree is preferable to a black-box model in clinical decision support.
\end{quote}
A short hospital-facing lecture cannot also walk through a derivation of Gini impurity, the split criterion used to grow a tree: that would add more technical machinery than the allotted time can carry.
The validator flags such a draft on cognitive load~\cite{sweller1988}, and the generator revises toward the interpretability trade-off above.

The generator sees $S$, $\pi$, $d$, optional extra instructions, and optional seed objectives.
It writes a prose statement using observable action verbs (explain, compare, diagnose, evaluate) and requires the target industry to appear.

A cross-model validator checks duration realism (a speaking-rate estimate against $d$), audience appropriateness (vocabulary and prior knowledge versus $\pi$, while preserving the seed's core aims), cognitive load (how many technical ideas need explaining in the allotted time), and clarity (concrete verbs).
Duration arithmetic is programmatic; the LLM judges audience fit, load, and specificity.

\subsection{Stage 2: Planning with Retrieved Domain Grounding}

This stage writes a lecture plan and retrieves an academic case study that matches the topic and the target industry.
Later stages read timing and visual intent from the plan, so parts stay consistent when they are generated in parallel.

In the running example, retrieval returns a published study on decision trees for 30-day hospital readmission in heart-failure patients.
The iris example is dropped.
The plan becomes a sequence of timed parts around that clinical setting; a draft part that stacked two dense visuals into a short segment is flagged as overload and simplified to a single feature-importance chart.

An academic search returns candidate papers.
A lightweight judge scores topic fit (same method as the seed) and industry fit (the application must be the target domain; a healthcare paper cannot satisfy a finance query).
Sources that fit are cited in the video; if none fit, planning continues.

Each part of the plan carries a title, an estimated duration, and a description of the intended visual.
Validation checks coherence of examples, one idea per part, and whether the intended visual can be rendered as a slide.
The validator can request regeneration of individual parts.

\subsection{Stage 3: Narration}

Stage~3 produces the spoken text for every part in one pass, so the lecture has a single narrative arc, a consistent register, and natural transitions.

In the running example, the seed says: ``The algorithm selects the feature that maximizes information gain. Consider the iris dataset\ldots''
The generated narration walks through 2{,}000 heart-failure patients (age, comorbidities, length of stay, 30-day readmission) and asks which feature and threshold best separates readmitted from non-readmitted patients.

The generator sees the full plan (and the retrieved paper when one was kept), $S$, and $\pi$.
A subsequent pass enforces a conversational register (Mayer's personalization principle).
Duration is steered with a speaking-rate estimate against each part's planned time.

\subsection{Stage 4: Parallel Visual Creation}

For each part, Stage~4 produces an HTML slide and, when the plan asks for one, a chart or diagram.
A visual-generation subagent runs on every part in parallel, in three steps: ideation, execution, and verification.

In the running example, part~3 becomes a two-column slide: bullets on Gini impurity beside a horizontal bar chart of feature importances (age, comorbidity index, length of stay, HbA1c) for the readmission model.
A vision check catches overlapping labels; the chart is regenerated with larger type.

Ideation, given the part plan and its narration, specifies text, whether a chart or diagram is needed, and layout.
Mayer's redundancy principle says learners process a lecture better when slides add structure instead of repeating the spoken words, so ideation keeps bullets short.
Execution writes the HTML markup and, in parallel, the code that renders charts and diagrams.
Verification first runs programmatic checks for missing images, math likely to overflow the slide, and content cropped at the edge, then a cross-model vision critique of sizing, readability, and fidelity to the ideation plan.
On failure, only execution is retried; the ideation plan is kept.

\subsection{Stage 5: Progressive Reveal and Assembly}

Learners follow a lecture more easily when each visual appears as it is mentioned (Mayer's signaling principle) and when related words and graphics sit together on the slide (spatial contiguity).
Stage~5 implements those principles by revealing slide elements as the narrator introduces them.

In the running example, the part~3 slide appears in four steps: the title; the first bullets as Gini is explained; the bar chart as the narrator turns to ``which features matter most for our readmission patients''; then the interpretation bullets.

Reveal timing is derived from the narration and checked programmatically.
Audio is then synthesized from the part transcripts and assembled with the reveal sequence into a single video.
That assembly step uses no LLM calls.

\section{Evaluation}\label{sec:method}

The study asks whether domain-matched experts judge Bespoke videos adequate, and how overall quality varies across topics, durations, and audiences.
Reviewers also scored how deeply industry-targeted videos felt personalized relative to generic-audience versions of the same material.

\subsection{Corpus}

We generated videos from 31 seed lectures drawn from a set of graduate courses in analytics, machine learning, and optimization offered in the same term at a research university.
Ten of those lectures (modules 4 and 8) were used while developing the pipeline; the other 21 are a held-out set.
Each seed was generated for three industries (healthcare, finance, energy) and a generic audience, at two durations: medium (5--15 minutes, key concepts) and long (matching the seed up to about 45 minutes; omitted when the seed was already short).
The pipeline produced over 200 videos.
No rated video was edited by hand after generation.
Twenty-five volunteers reviewed a 92-video subset (healthcare 32, generic 31, finance 22, energy 7; medium 58, long 34), one review per video.
The subset is the largest set we could cover under each reviewer's time budget, matched to the topics they listed as within their expertise.

Generation used Anthropic's Claude Sonnet~4.6 API and OpenAI's GPT-5.2 API, with Sonnet generating and GPT-5.2 validating, and a bound of three refinement iterations.
Audio was synthesized and assembled with the slides as described in Stage~5.

\subsection{Rubric}

Reviewers scored four dimensions, Content (A), Personalization (B), Pedagogical effectiveness (C), and Production (D), plus a global item $G$ and a confidence item.
Each item uses a five-point scale ($i$--$v$, mapped to 1--5).
The form gives a written description of a failing lecture ($i$), an acceptable lecture ($iii$), and an exemplary lecture ($v$); $ii$ and $iv$ sit between those anchors.
The full wording is in Appendix~\ref{app:rubric}, which reproduces the instrument as reviewers saw it.

\textbf{Content (A).}
A has two sub-items: whether claims are accurate (A1), and whether the lecture stays within the stated objectives and duration (A2).
The overall A score of $iii$ means ``acceptable to show to students''; $v$ means ``would pass peer review as a teaching resource.''
Reviewers already knew the topic and watched without the seed, so A records errors they noticed while watching.

\textbf{Personalization (B).}
B has two sub-items.
B1 is cumulative: industry vocabulary is required before professional motivation, domain interpretation, and connection to existing expertise.
B2 asks whether cognitive level, jargon, and assumed knowledge match the stated audience.
Overall B ranges from ``could be shown to any audience'' to ``feels authored by someone who understands this audience's daily work.''

\textbf{Pedagogical effectiveness (C).}
C has four sub-items: narrative coherence (C1); teaching strategies such as analogies, worked examples, and simple-to-complex progression (C2); visual--concept alignment (C3); slide layout as cognitive design (C4).
Overall C ranges from ``a raw encyclopedia article read aloud'' to ``the concept feels obvious by the end.''

\textbf{Production (D).}
D has two sub-items: spoken language quality (D1) and slide visual quality (D2).
Overall D ranges from ``would not be shown in any professional context'' to ``production quality actively helps.''

\textbf{Global quality (G).}
$G$ is a holistic score given independently of A--D.
The $iii$ anchor is ``comparable to a standard MOOC lecture,'' a midpoint written for reviewers who teach or TA MOOCs and analytics courses.
The review set did not include seed lectures or published MOOCs.

Each dimension and G had optional free text; G required a written comment.
A lecture can be factually correct and generic (high A, low B), or pedagogically strong and visually poor (high C, low D).

\subsection{Expert Panel}

Twenty-five reviewers scored only videos in their domain.
They include MOOC instructors, teaching assistants for analytics and machine learning courses, and doctoral researchers whose work includes applications in healthcare, finance, or energy.

We assigned videos by solving an optimal matching problem: each video goes to a reviewer who listed all of that video's topics, each video to at most one person, maximizing minutes reviewed under a cap of 120 minutes and 6 videos.
We solve that matching with a mixed-integer program.
Few reviewers listed energy, so that cell is small ($n=7$).
Each video is seen by exactly one domain-matched reviewer ($k=1$).
We chose $k=1$ so the panel could cover more of the corpus and test generalizability across lectures, rather than collecting multiple ratings of the same video.
Reviewers did not see other variants of the same topic or the seed, so a score reflects a single viewing rather than a side-by-side comparison.
Each video is scored by one reviewer, so we have no inter-rater agreement statistic.
Reviewers score several videos each (mean 3.7).
A random-intercept model on $G$ attributes about 35\% of residual variance to the reviewer (ICC $= 0.35$).
We therefore report reviewer-clustered intervals; the corresponding design effect is about 1.9.
We report a mixed model for the B1 contrast.
Matching by expertise also means that scores for an industry come from that industry's reviewers.

We report means and standard deviations by domain and duration, and the share of videos with $G\geq 3$.
Reviewers also recorded confidence (mean $3.33$ on the same 1--5 scale).
Free-text comments were grouped into recurring strengths and failures (voice, reveal lag, layout, pedagogy).

\paragraph{Ethics.}
Reviewers were recruited through the university's MOOC service and through the department of one of the authors, and participated out of professional interest.
Several are themselves MOOC instructors.
They volunteered unpaid.
The assignment cap was 120 minutes and 6 videos.
None of the reviewers authored or TA'd the seed lectures they scored.
They consented to publication of their written comments.
We did not seek institutional review: the task was watching lecture videos and scoring them, and our institution did not require IRB review for this study.
The seed-lecture instructors consented to use of the transcripts.

\begin{figure*}[t]
  \centering
  \includegraphics[width=\linewidth]{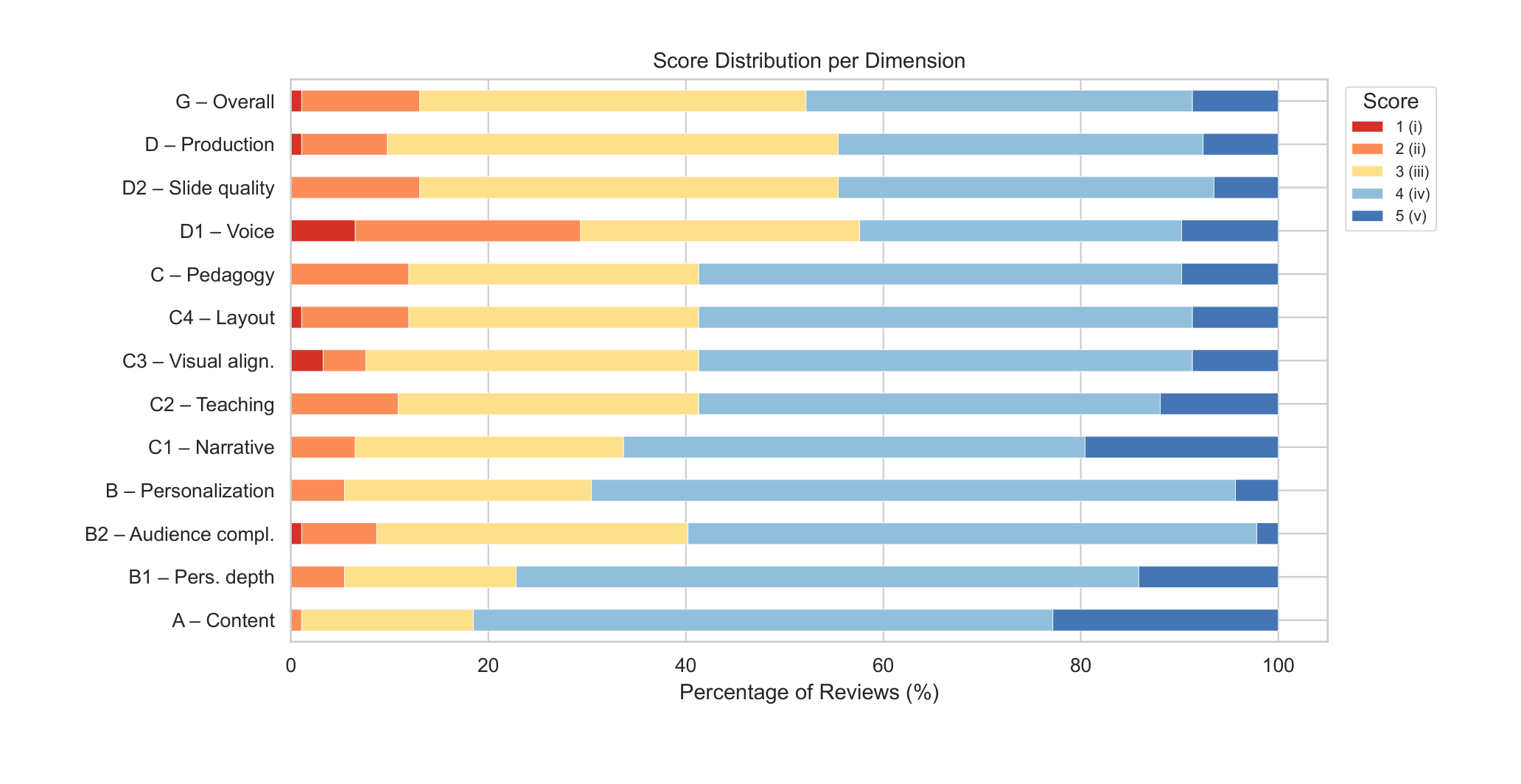}
  \Description{Stacked bars of expert scores from 1 to 5 on content, personalization, pedagogy, production, and global quality. Most mass is at 3 or above. Content is strongest; voice is weakest.}
  \caption{Experts judged 87\% of the videos comparable to a standard MOOC lecture on overall quality ($G$; midpoint $iii=3$), and 48\% above that bar. On factual content ($A$), 99\% were judged sufficient: claims the reviewers found sound enough to show to students.}
  \label{fig:scores}
\end{figure*}

\section{Results}\label{sec:results}

\subsection{Overall Quality}

Figure~\ref{fig:scores} summarizes the 92 reviews.
Reviewers judged 87\% of videos at or above ``comparable to a standard MOOC lecture'' ($G\geq 3$; reviewer-clustered bootstrap 95\% CI $[76\%, 96\%]$).
Mean global quality is $G=3.42$ out of $5$ (SD~$=0.85$); 48\% scored $G\geq 4$ (CI $[33\%, 63\%]$), above that midpoint anchor.
A reviewer random-intercept model puts mean $G$ at $3.39$.

Content is the strongest dimension ($A=4.03$, SD~$=0.67$): 99\% score $\geq 3$ and 82\% score $\geq 4$.
Personalization depth $B1=3.86$ (94.6\% $\geq 3$); audience calibration $B2=3.52$.
Pedagogy $C=3.57$, led by narrative coherence ($C1=3.79$).
Production is weakest ($D=3.41$), driven by voice ($D1=3.16$, SD~$=1.09$).

Generic-audience videos still average $B1=3.65$.
Those videos are already rewritten for working professionals: shorter than the seed, in a conversational register, with concrete examples in place of the original academic framing.
Industry-targeted videos score 0.32 points higher on B1 (3.97 vs.\ 3.65).
A mixed model with a reviewer random intercept estimates that increment at 0.25 (95\% CI $[-0.06, 0.56]$, $p=0.11$).

\subsection{Association between Dimensions and Global Score}

\begin{figure}[t]
  \centering
  \includegraphics[width=\linewidth]{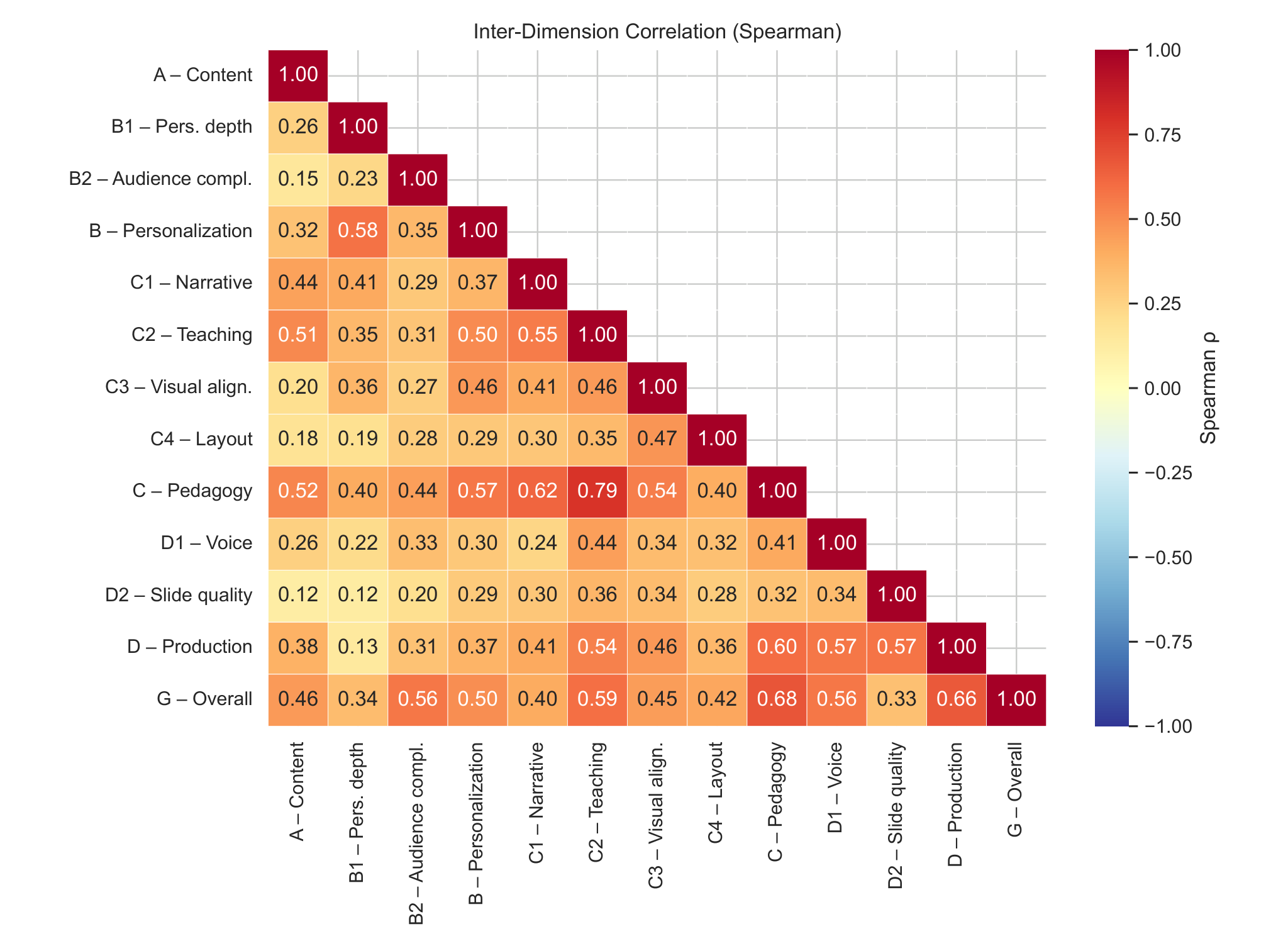}
  \Description{Lower-triangular Spearman correlation heatmap among rubric items. Pedagogy and production associate most strongly with global quality. Content is more independent.}
  \caption{Spearman correlations among rubric items. Pedagogy (C) and production (D) associate most strongly with global quality (G). Content (A) is relatively independent.}
  \label{fig:corr}
\end{figure}

Pedagogy and production both associate with $G$ (Spearman $\rho=0.68$ between C and G; $\rho=0.66$ between D and G), which matches the free-text emphasis on pacing, reveal timing, and voice.
Content is weaker ($\rho=0.46$): with 99\% of videos already $\geq 3$ on A, accuracy has little leftover variance.
Figure~\ref{fig:corr} shows B, C, and D intercorrelated ($\rho=0.37$--$0.60$); A is more independent of B and D ($\rho=0.32$--$0.38$) and still correlates with C ($\rho=0.52$).
Voice has the lowest mean (3.16) and the highest share below 3 (29\%).

\subsection{Illustrative Reviewer Comments}

Reviewers left 235 free-text comments on 73 of 92 videos.
Slide visual quality, slide layout, and voice drew the most remarks (31, 30, and 27 comments).
Comments also address personalization, pacing, and whether the lecture works as teaching material.

\textbf{Voice.}
Twenty-seven comments concern spoken delivery, mostly pronunciation.
Several videos read two-digit numbers digit by digit (``30'' as ``three zero,'' ``60'' as ``six zero''), in at least six cases ``repeated across the entire video'' and ``very distracting.''
Reviewers also flag missing pauses (``He talks far too fast \ldots I miss pauses, especially after he asks a question'') and a playback speed that ``sounds like a 1.3 playback.''
Several still judge the voice usable: ``Definitely can tell it is a robot, but not distracting at all''; ``The voice is nice to listen to.''

\textbf{Slide timing and layout.}
Comments on layout often describe content spoken before it appears: ``the narration dictates the formula before displaying it on the slide''; ``it talks about equations without displaying the equation and the delay in display is very large.''
When progressive reveal works, reviewers notice: ``Slide building helps a lot here, because it makes the slides easier to follow while the speaker is explaining the concepts.''
Recurring visual issues include unused space, overlapping labels, cropped plots, and text-heavy slides: ``the amount of unused space on some slides, which can be distracting''; ``There could be less text overall.''

\textbf{Pedagogy and personalization.}
Some lectures start abruptly (``I was thrown into the lecture without a general overview'') or stay at a flat pace with few pauses for the learner to think.
Industry versions are often recognized as targeted (``perfectly tailored to an audience with that background''; ``Referred to how clinicians explain their decisions in real life'').
Generic-audience videos can still pick up an industry flavor from an example (``Very healthcare-focused, more appropriate for healthcare professionals than for general working professionals'').

\textbf{Reception.}
Comments on the lecture as teaching material are often positive: ``This is a very good quality lecture. I would pay to watch this lecture''; ``Good pace of information, lots of helpful comments/examples''; ``Good intuition comments about what overfitting is.''
A few reviewers argue that slides plus synthetic voice have a ceiling: ``it lacks the human engagement piece that voiceover restricts you to''; another wrote that the delivery ``lacks soul.''
Several asked for more active learning (``a pause to ask the learner to be active'') or denser visuals.

\subsection{Scores by Duration, Audience, Held-Out Set, and API Cost}

Mean $G$ ranges from 3.32 on generic-audience videos to 3.55 on finance videos (Table~\ref{tab:domain}). Reviewers scored only videos in their domain, so the table is descriptive.

Medium and long videos are similarly close (3.36 vs.\ 3.53; Mann--Whitney $U$ $p=0.60$; difference $0.17$, 95\% CI $[-0.17, 0.50]$; Table~\ref{tab:depth}).
Those two settings differ in duration and in how much of the seed they cover (key concepts versus matching the seed).

Videos from the 21 held-out lectures score $G=3.36$, versus $3.51$ on the 10 development lectures (difference $-0.15$, 95\% CI $[-0.48, 0.18]$).
Taken together, overall quality does not drop sharply on new lectures, other industries, or the longer format.

Mean API cost is about \$0.22 per minute, or about \$5 for a typical video.

\begin{table}[t]
  \caption{Mean scores by target domain (SD in parentheses). Reviewers scored only videos in their domain, so these rows are descriptive.}
  \label{tab:domain}
  \resizebox{\columnwidth}{!}{%
  \begin{tabular}{@{}lrrrrrr@{}}
    \toprule
    Domain & $n$ & A & B & C & D & G \\
    \midrule
    Energy & 7 & 4.29 (0.76) & 3.71 (0.76) & 3.86 (1.22) & 3.71 (0.79) & 3.43 (0.79) \\
    Finance & 22 & 3.91 (0.75) & 3.68 (0.65) & 3.77 (0.75) & 3.59 (0.73) & 3.55 (0.86) \\
    Healthcare & 32 & 4.09 (0.69) & 3.81 (0.59) & 3.50 (0.84) & 3.38 (0.83) & 3.44 (0.88) \\
    Generic & 31 & 4.00 (0.58) & 3.55 (0.68) & 3.42 (0.77) & 3.26 (0.73) & 3.32 (0.87) \\
    \bottomrule
  \end{tabular}}
\end{table}

\begin{table}[t]
  \caption{Mean scores by duration (SD in parentheses).}
  \label{tab:depth}
  \resizebox{\columnwidth}{!}{%
  \begin{tabular}{@{}lrrrrrr@{}}
    \toprule
    Duration & $n$ & A & B & C & D & G \\
    \midrule
    Medium & 58 & 4.00 (0.65) & 3.60 (0.67) & 3.50 (0.80) & 3.41 (0.80) & 3.36 (0.91) \\
    Long & 34 & 4.09 (0.71) & 3.82 (0.58) & 3.68 (0.88) & 3.41 (0.82) & 3.53 (0.75) \\
    \bottomrule
  \end{tabular}}
\end{table}

\section{Discussion}\label{sec:discussion}

\subsection{Implications for Instructors and L\&D}

Bespoke produces a first cut of a professional-audience lecture at about \$0.22 API cost per minute.
For an instructor or learning team who already has a seed lecture, videos in this sample often meet the rubric's MOOC-comparable bar.
The remaining human pass is voice timing, slide--narration sync, and slide rendering.
Those issues are easy to catch in a short skim.

\begin{sloppypar}
Audience calibration (B2) trails personalization depth (B1).
Putting hospital vocabulary on a decision tree is easier than deciding how much Gini impurity a clinical audience needs.
\end{sloppypar}
A practitioner panel would be a stricter test of that calibration.

Content scores are high ($A=4.03$; 99\% $\geq 3$).
Reviewers rarely flagged factual errors while watching; a later audit against the seed would still be useful before a high-stakes deployment.

\subsection{Implications for Platforms}

Recommendation chooses which existing video a learner sees.
In-video Q\&A (SAM, Generative Lecture) adds interaction on top of a fixed recording.
Audience-level generation is a third lever: new slides, narration, and examples for a stated professional group.
Platforms that already host one master lecture per topic could treat industry variants as generated siblings.

Authoring tools such as VIVID and PedaCo-Gen keep the instructor in the loop.
A natural next system would expose the compiled artifacts (objectives, plan, slides, transcript) for a short instructor pass before learners see the video.
The qualitative results say where that pass should look: voice, reveal lag, and slide rendering.

If a platform generated industry variants at catalog scale, those videos would inherit the seed instructor's copyright and any citation obligations from retrieved case studies.
Consent and attribution for the seed, and for papers the pipeline quotes, would have to be in place before variants follow paying audiences (a hospital L\&D catalog versus a public MOOC).

\subsection{Limits}

This paper reports expert judgments of video quality.
A learner study of learning, transfer, completion, or engagement is future work.
Raters are instructors, TAs, and doctoral researchers with industry applications; each video is seen once ($k=1$); there is no inter-rater reliability estimate; and the ``standard MOOC'' phrase is a rubric midpoint.
The extra B1 increment for a named industry, beyond a generic professional rewrite, is small and not resolved under reviewer clustering.

\section{Conclusion}

Bespoke takes the transcript of an existing lecture and generates a new video, with slides, narration, and charts, customized to a stated professional audience.
A five-stage workflow compiles those artifacts and revises them across two model families.

On a 92-video expert sample, reviewers judged 87\% at or above the rubric midpoint corresponding to ``a standard MOOC lecture's quality'' (mean $G=3.42$ out of $5$).
Overall quality was similar across industries, durations, and the 21-lecture held-out set, at about \$0.22 API cost per minute.
Voice timing, slide--narration sync, and slide rendering are the main remaining issues.

A next step is to measure whether these videos help professionals learn.

\bibliographystyle{ACM-Reference-Format}
\bibliography{bespoke-chi2027}

\appendix

\section{Evaluation Rubric}\label{app:rubric}

Experts watched each assigned video and rated it relative to the stated learner profile, learning objectives, and target duration.
Every item uses a five-level scale ($i$--$v$) with behavioral anchors at $i$, $iii$, and $v$:
$i$ clearly fails;
$ii$ below expectations;
$iii$ meets a basic acceptable standard;
$iv$ above expectations, short of exceptional;
$v$ exemplary.
When in doubt, reviewers scored what they observed.
Each dimension also has an overall item that need not equal the item-level scores.
G requires a written comment; other free-text fields are optional.

\begin{shaded}
\small
\raggedright
\newcommand{\ranchor}[2]{\par\noindent\hangindent=1.6em\hangafter=1\textbf{#1.}\ #2\par}

\noindent\textit{Reviewer instrument (verbatim).}

\medskip
\noindent\textbf{A. Content}

\noindent\textit{A overall.}
\ranchor{$i$}{serious errors or scope failures; a hallucinated or heavily fabricated summary.}
\ranchor{$iii$}{acceptable to show to students; minor inaccuracies that do not undermine the main message; non-peer-reviewed draft notes.}
\ranchor{$v$}{every claim correct and verifiable; would pass peer review as a teaching resource.}

\medskip
\noindent\textbf{B. Personalization}

\noindent\textit{B1 depth (cumulative).}
\ranchor{$i$}{no personalization; generic or wrong-domain examples.}
\ranchor{$ii$}{surface only (industry vocabulary; examples could belong to any domain).}
\ranchor{$iii$}{partial; the application may not match this learner's motivations.}
\ranchor{$iv$}{mostly deep: professional motivation, domain interpretation, at least one connection to existing expertise.}
\ranchor{$v$}{fully deep and relevant; a practitioner would recognize it as pertinent to their work.}

\noindent\textit{B2 complexity.}
\ranchor{$i$}{drastically too advanced or too elementary.}
\ranchor{$iii$}{broadly appropriate, with occasional missteps.}
\ranchor{$v$}{pitched at the boundary of what this audience already knows.}

\noindent\textit{B overall.}
\ranchor{$i$}{could be shown to any audience.}
\ranchor{$iii$}{some genuine tailoring; a general-purpose MOOC with optional applied tracks.}
\ranchor{$v$}{feels authored by someone who understands this audience's daily work.}

\medskip
\noindent\textbf{C. Pedagogical effectiveness}

\noindent\textit{C1 narrative.}
\ranchor{$i$}{disconnected parts; could be reordered without loss.}
\ranchor{$iii$}{followable progression, some loose or redundant parts.}
\ranchor{$v$}{a single driving argument; a student could reconstruct the outline.}

\noindent\textit{C2 teaching strategies.}
\ranchor{$i$}{flat fact delivery.}
\ranchor{$iii$}{multiple strategies (analogies, worked examples, simple-to-complex) applied inconsistently.}
\ranchor{$v$}{varied and layered strategies; a novice finishes with understanding, not only exposure.}

\noindent\textit{C3 visual--concept alignment.}
\ranchor{$i$}{visuals irrelevant, missing, or confusing.}
\ranchor{$iii$}{most visuals support the narration.}
\ranchor{$v$}{every visual is the best choice for the point it explains.}

\noindent\textit{C4 layout (not aesthetics).}
\ranchor{$i$}{density and missing hierarchy hinder processing.}
\ranchor{$iii$}{readable and organized; hierarchy present but not exploited.}
\ranchor{$v$}{hierarchy and signaling make the argument visible; nothing extra on the slide.}

\noindent\textit{C overall.}
\ranchor{$i$}{a raw encyclopedia article read aloud.}
\ranchor{$iii$}{a decent MOOC lecture.}
\ranchor{$v$}{the concept feels obvious by the end.}

\medskip
\noindent\textbf{D. Production}

\noindent\textit{D1 language and voice} (weaker of script vs.\ voice).
\ranchor{$i$}{artifacts that harm comprehension.}
\ranchor{$iii$}{acceptable but inconsistent; a decent MOOC narration.}
\ranchor{$v$}{written to be heard; sounds like a fluent speaker.}

\noindent\textit{D2 slide visual quality.}
\ranchor{$i$}{broken or unusable (overlap, crop, broken formulas).}
\ranchor{$iii$}{usable; few minor imperfections.}
\ranchor{$v$}{flawlessly rendered; deliberate visual language.}

\noindent\textit{D overall.}
\ranchor{$i$}{would not be shown in any professional context.}
\ranchor{$iii$}{clean enough for a standard online course.}
\ranchor{$v$}{production quality actively helps.}

\medskip
\noindent\textbf{G. Global quality}

Considering content, personalization, teaching, and production for the stated learner:
\ranchor{$i$}{would actively harm understanding.}
\ranchor{$iii$}{an acceptable lecture; the learner would reasonably meet the objectives. Reference level: ``a standard MOOC lecture.''}
\ranchor{$v$}{the learner would recommend it to a colleague in the same role.}

\noindent\textit{Confidence.}
\ranchor{$i$}{not familiar with the topic or audience.}
\ranchor{$v$}{expert in this topic and familiar with the target audience.}
\end{shaded}

\end{document}